\documentclass[12pt]{article}

\usepackage[paper=a4paper, margin=2cm]{geometry}
\usepackage[russian,english]{babel}
\usepackage[cp1251]{inputenc}
\usepackage{amsmath}

\usepackage[logos]{amsbib}

\begin{document}
%%%%%%%%%%%%%%%%%%%%%%%%%%%%%%%%%%%%%%%%%%%%%%%%%%%%%%%%%%%%%%%%%%%%%%%%%%%%%%%%%%%%

\date{}
\renewcommand{\refname}{References}

\author{A.~Yu.~Samarin%, samarinay@yahoo.com
}
\title{Quantum Particle Motion in Physical Space}

\maketitle

\centerline{Samara Technical State University, 443100 Samara, Russia}
\centerline{}

\abstract{  Using Feynman's representation of the quantum evolution and considering a quantum particle as a matter field (continuous medium), it is shown that individual particles of the field have unique paths of the motion. This allows describing motion of the quantum particle continuous medium by Lagrange's method. It is shown that form of the real individual particle path is determined by classical minimum action principle.

{
{\bf Keywords:} quantum evolution in physical space, matter field, continuous medium, mechanical motion, unobservable motion characteristics.
}

%%%%%%%%%%%%%%%%%%%%%%%%%%%%%%%%%%%%%%%%%%%%%%%%%%%%%%%%%%%%%%%%%%%%%%%%%%%%%%%%%%%%

\begin {center} \textbf{1. INTRODUTION}\\
\end {center}

 If the wavefunction expresses properties of a matter field
 \footnote{For correctness of this assumption the wavefunction has to be nondegenerate. This requirement does not lead to a loss of consideration generality because of superposition principle. In the case of a degenerate wavefunction the quantum particle is represented as a continua
collection, every element of which corresponds to a nondegenerate wavefunction.}, then the reduction process is real nonlocal transformation of the continuous medium. In the paper ~\cite{bib:L} the wavefunction reduction has been described as such real process. Suppose the quantum particle is a matter field in physical space.  Then, the set of physical space coordinates that are applicable domain of the wavefunction has cardinality of continuum. This means that mechanical motion of the quantum particle continuous medium cannot be described by observables (any collection of observables values is countable ~\cite{bib:F}).

Evolution of the wavefunction in physical space is considered in Feynman's interpretation of quantum mechanics ~\cite{bib:KV, bib:V}. Note that there is no necessity to use the observables for evolution description in this interpretation per se and it can be considered as the Eulerian method of continuum motion description. It remains to check that the Lagrangian method exists. This is the main goal of the paper.

Consider one-dimensional motion of the quantum continuum.  Denote by $\Psi_{t_{1}}(x_{1}) $  and $\Psi_{t_{2}}(x_{2}) $ wavefunctions of the initial and final states, where $x_{1} $, $x_{2} $  are the initial and final space coordinates, $t_{1} $, $t_{2} $ are the respective instants of time. The dynamic low in Feynman theory has the form of the integral wave equation:

 \begin{equation}\label{eq:math:ex1}
    \Psi_{t_{2}}(x_{2})
    =\idotsint K_{t_{2},t_{1}}(x_{2}, x_{1})\Psi_{t_{1}}(x_{1})\,dx_1.
\end{equation}
Kernel of the integral evolution operator $K_{t_{2},t_{1}}(x_{2},x_{1}) $   considered as function of the space variables has parametrical time dependence (as the wavefunctions ~\cite{bib:F})). It is the transition amplitude that has the form of a path integral:
\begin{equation}\label{eq:math:ex2}
    K_{t_{2},t_{1}}(x_{2},x_{1})=\int [dx(\tau)]\exp{\frac{i}{\hbar}S_{1,2}[x(\tau)]}.
\end{equation}
The action functional $S[x(\tau)] $  is the time integral of the Lagrangian taken along the virtual path $x(\tau) $. The integration is to be taken over all virtual paths $x(\tau)] $; it has the form of a continual integral ~\cite{bib:PRL}. The transition amplitudes for different values of the initial space coordinates are independent on each other. This allows considering these transitions as the result of independent mechanical motion of the individual particles (the individual particles are the continuum elements such that there is biunique correspondence between these elements and coordinates of the space occupied by the continuous medium\footnote{It is possible for nondegenerate wavefunctions. }). Suppose that the individual particles conserve identity in the transition process \eqref{eq:math:ex1}. Then the physical quantities have to exist such that the mechanical motion is described by these quantities at any transition time. These quantities have to describe the motion state of the individual particle at any instant of transition time. Generalizing \eqref{eq:math:ex2}, we get for the motion characteristics (the transition quantities) following expression ~\cite{bib:V}:
\begin{equation}\label{eq:math:ex3}
    \langle f(\tau)\rangle_{1,2}=\int [dx(\tau)]f_{x(\tau)}(\tau)\exp{\frac{i}{\hbar}S_{1,2}[x(\tau)]},
\end{equation}
where $f_{x(\tau)}(\tau) $ is classical kinematic quantity for the virtual path $x(\tau) $. If to attribute these quantities to the real paths of the individual particles, it remains to check that a unique path exists.

\begin {center} \textbf{\textbf{2. THE REAL PATH}}\\
\end {center}
Following theorem is useful for this goal.
\centerline{}
\centerline{}
{\bf Theorem}.
\emph{If the physical quantity $f=f(x) $ is function of the individual particle coordinate $x $ for each virtual path, then the unique path $x_{m}(\tau) $ exists such that the transition quantity $\langle f\rangle $ is determined by this path.}
\centerline{}
\centerline{}
{\bf Proof.} Let the classical physical quantity be the function of the individual particle coordinate: $f=f(x) $. Suppose the coordinate values are attributes of the path $\gamma\equiv x(\tau) $, then values of the quantity $f $ is attribute of the same virtual path too: $f_{\gamma}=f(x_{\gamma}) $ \footnote{Here and further we suppose that the quantum particle moves in stationary potential field.}. Using expression \eqref{eq:math:ex3}, we have
\begin{equation*}
    \langle f(\tau)\rangle=\int [d\gamma]f(x_{\gamma})\exp{\frac{i}{\hbar}S[\gamma]},
\end{equation*}
Suppose $\{x_{\gamma}\} $ is set of the variables having continuum cardinal numbers. Then the quantity $\langle f\rangle $ can be considered as the function of these variables. If to expand it into a power series, then terms of the expansion will be continual integrals. For the linear term we have:
\begin{equation*}
   \int\biggl( \frac{\partial}{\partial x_{\gamma'}}\int [d\gamma]f(x_{\gamma})\exp{\frac{i}{\hbar}S[\gamma]}\biggl)^{(0)}x_{\gamma'}[d\gamma'].
   \end{equation*}
For the term corresponding to the quadratic form, we obtain
\begin{equation*}
   \frac{1}{2}\int x_{\gamma'}[d\gamma']\int x_{\gamma''}\biggl( \frac{\partial^{2}}{\partial x_{\gamma'}\partial x_{\gamma''} }\int [d\gamma]f(x_{\gamma})\exp{\frac{i}{\hbar}S[\gamma]}\biggl)^{(0)}[d\gamma''].
\end{equation*}
Since the expansion terms, corresponding to the different variables, are taken into account in the path integral doubly, then the corresponding terms have to be divided by 2 (the terms corresponding to the same variables have to be divided by 2 in line with usual for the Taylor expansion reason). Other terms of the expansion have the same structure. Therefore
\begin{multline*}
 \biggl(\int [d\gamma]f(x_{\gamma})\exp{\frac{i}{\hbar}S[\gamma]}\biggl)^{(0)}+\int x_{\gamma'}\biggl( \frac{\partial}{\partial x_{\gamma'}}\int [d\gamma]f(x_{\gamma})\exp{\frac{i}{\hbar}S[\gamma]}\biggl)^{(0)}[d\gamma']+\\
    +\frac{1}{2}\int x_{\gamma'}[d\gamma']\int x_{\gamma''}\biggl( \frac{\partial^{2}}{\partial x_{\gamma'}\partial x_{\gamma''} }\int [d\gamma]f(x_{\gamma})\exp{\frac{i}{\hbar}S[\gamma]}\biggl)^{(0)}[d\gamma'']+...+.
\end{multline*}
Substituting the derivative $\frac{\partial}{\partial x\gamma'}$   by $\frac{\partial}{\partial\langle x\rangle}$ in the last expression, we get
\begin{multline}\label{eq:math:ex4}
 \langle f\rangle=\biggl( \int [d\gamma]f(x_{\gamma})\exp{\frac{i}{\hbar}S[\gamma]}\biggl)^{(0)}+\\ +\biggl( \frac{\partial}{\partial \langle x\rangle}\int x_{\gamma'} [d\gamma]f(x_{\gamma})\exp{\frac{i}{\hbar}S[\gamma]}\biggl)^{(0)}\int x_{\gamma'}\biggl(\frac{\partial\langle x\rangle}{\partial x_{\gamma'}}\biggl)^{(0)}[d\gamma']+\\
    +\biggl( \frac{1}{2}\frac{\partial^{2}}{\partial \langle x\rangle\partial \langle x\rangle} \int [d\gamma]f(x_{\gamma})\exp{\frac{i}{\hbar}S[\gamma]}\biggl)^{(0)}\times\\\times \int x_{\gamma'}\biggl(\frac{\partial\langle x\rangle}{\partial x_{\gamma'}}\biggl)^{(0)}[d\gamma']\int x_{\gamma''}\biggl(\frac{\partial\langle x\rangle}{\partial x_{\gamma''}}\biggl)^{(0)}[d\gamma'']+...+.
\end{multline}

Every collection of the kinematics characteristics corresponds to a unique virtual path. In order to express this mathematically, we can introduce the following object:
\begin{equation*}
\delta \bigl[\gamma'-\gamma\bigl]=\left \{ \begin{aligned}
  0 \quad\text{when}\quad\gamma'\neq \gamma,\\
  1 \quad\text{when}\quad \gamma'=\gamma.
\end{aligned} \right.
\end{equation*}
Taking into account the physical sense of this functional and the formal similarity with Dirac's  $\delta $-function we have
\begin{equation*}
  \begin{aligned}
  &\int\delta[\gamma'-\gamma]f[\gamma'][d\gamma']=f[\gamma],\\
  &\int\delta[\gamma'-\gamma][d\gamma']=1.
\end{aligned}
\end{equation*}
It now follows that
\begin{equation*}
 \frac{\partial \langle f\rangle}{\partial x_{\gamma}}=\frac{\partial }{\partial x_{\gamma}}\int f(x_{\gamma'})\exp\frac{i}{\hbar}S_{\gamma'}[d\gamma']=\int\delta[\gamma-\gamma']\exp\frac{i}{\hbar}S_{\gamma'} \frac{\partial f(x_{\gamma'})}{\partial x_{\gamma'}}[d\gamma'].
\end{equation*}
Using \eqref{eq:math:ex3} for the transition quantity $x $, we get
\begin{equation*}
 \langle x\rangle=\int[d\gamma]x_{\gamma}\exp\frac{i}{\hbar}S[\gamma].
\end{equation*}
Then
\begin{multline*}
   \int x_{\gamma'}\biggl( \frac{\partial \langle x\rangle}{\partial x_{\gamma'}} \biggl)^{(0)}[d\gamma']=\int x_{\gamma'}\biggl( \frac{\partial}{\partial x_{\gamma'}} \int  x_{\gamma}\exp\frac{i}{\hbar}S_{\gamma}[d\gamma]\biggl)^{(0)}[d\gamma']=\\
    =\int x_{\gamma'}\biggl(\int\delta[\gamma-\gamma']\exp\frac{i}{\hbar}S_{\gamma}[d\gamma]\biggl)^{(0)}[d\gamma']= \int x_{\gamma'}\exp\frac{i}{\hbar}S_{\gamma'}[d\gamma']=\langle x\rangle. \end{multline*}
    Combining these results and expansion \eqref{eq:math:ex4}, we obtain
\begin{equation}\label{eq:math:ex5}
  \langle f\rangle=\langle f\rangle_{0}+\biggl(\frac{\partial\langle f\rangle}{\partial\langle x\rangle}\biggl)_{0}\langle x\rangle+\frac{1}{2}\biggl(\frac{\partial^{2}\langle f\rangle}{\partial\langle x\rangle^{2}}\biggl)_{0}\langle x\rangle^{2}+...+.
\end{equation}
Since the path integral is over all virtual path then there is the virtual path   such that
\begin{equation*}
  x_{m}(t)\equiv|\langle x(\tau)\rangle|.
\end{equation*}
In this case
\begin{equation*}
  \langle x(\tau)\rangle=x_{m}(\tau)\exp\frac{i}{\hbar}S_{m}[x_{m}(\tau)]+\int\limits_{\gamma\neq\gamma_{m}}[d\gamma]x_{\gamma}(\tau)\exp\frac{i}{\hbar}S[\gamma]\equiv|\langle x(\tau)\rangle|\exp i\varphi,
\end{equation*}
where the integral in last equation is over all virtual paths, excepting the path $\gamma_{m}$; $\varphi$ --- is phase of the transition coordinate. Since this expression must hold identically for any time and for any full set of the virtual paths, it follows that $\varphi=S[x_{m}(\tau)] $, and
\begin{equation*}
 \int\limits_{\gamma\neq\gamma_{m}}[d\gamma]x_{\gamma}(\tau)\exp\frac{i}{\hbar}S[\gamma]\equiv 0.
\end{equation*}
Therefore
\begin{equation}\label{eq:math:ex6}
  \langle x(\tau)\rangle=x_{m}(\tau)\exp\frac{i}{\hbar}S_{m}[x_{m}(\tau)].
\end{equation}
Combining \eqref{eq:math:ex5} and \eqref{eq:math:ex6}, we obtain
\begin{equation*}
  \langle f\rangle=\langle f\rangle^{(0)}+\biggl(\frac{\partial\langle f\rangle}{\partial x_{m}}\biggl)^{(0)} x_{m}+\frac{1}{2}\biggl(\frac{\partial^{2}\langle f\rangle}{\partial x_{m}^{2}}\biggl)^{(0)} x_{m}^{2}+...+.
  \end{equation*}
Since
\begin{multline*}
  \frac{\partial}{\partial x_{m}}\langle f\rangle=\int[d\gamma] \frac{\partial}{\partial x_{m}} f(x_{\gamma})\exp\frac{i}{\hbar}S_{\gamma}=\\
  =\int [d\gamma]\delta[\gamma-\gamma_{m}]\frac{\partial}{\partial x_\gamma} f(x_{\gamma})\exp\frac{i}{\hbar}S_{\gamma}=\frac{\partial}{\partial x_m} f(x_{m})\exp\frac{i}{\hbar}S[x_{m}(\tau)].
  \end{multline*}
  then
  \begin{equation*}
  \langle f\rangle=\biggl(f(x_{m})^{(0)}+\biggl(\frac{\partial f(x_{m})}{\partial x_{m}}\biggl)^{(0)} x_{m}+\frac{1}{2}\biggl(\frac{\partial^{2} f(x_{m})}{\partial x_{m}^{2}}\biggl)^{(0)} x_{m}^{2}+...+\biggl)\exp\frac{i}{\hbar}S[x_{m}(\tau)].
  \end{equation*}
Therefore
\begin{equation}\label{eq:math:ex7}
    \langle f(x)\rangle=f( x_{m})\exp\frac{i}{\hbar}S[x_{m}(\tau)]
\end{equation}

Thus, transition quantities for any individual particle are determined by the unique virtual path $x_{m}(\tau)$. Such situation corresponds to Lagrange's description of the continuum motion and the path $x_{m}(\tau)$ can be considered as a real path. In order to realize this description it is necessary to find the dynamic law that determines form of the real path.

\begin {center} \textbf{\textbf{3. THE DYNAMIC LAW}}\\
\end {center}

Let the individual particle moves from the point with the coordinate $x_{1}$ of physical space at the instant of time $t_{1}$  to the point with the coordinate $x_{1} $ of physical space at the instant of time $t_{2}$. Consider all virtual paths for this motion. Let $\delta\gamma$ be variation of the virtual paths path such that
\begin{equation*}
    \delta\gamma|_{\tau=t_{1}}=\delta\gamma|_{\tau=t_{2}}=0
\end{equation*}
Variations of the different virtual path are the same. Since the virtual paths set are exhaustive, we see that
\begin{equation*}
   \int[d\gamma]\exp\frac{i}{\hbar}S[\gamma]= \int[d(\gamma+d\gamma)]\exp\frac{i}{\hbar}S[\gamma+d\gamma]
\end{equation*}
Expanding into the series the complex exponent in the second integral, we obtain
\begin{equation*}
  \exp\frac{i}{\hbar}S[\gamma+d\gamma]=\exp\frac{i}{\hbar}S[\gamma]+\frac{i}{\hbar}\delta S[\gamma]\exp\frac{i}{\hbar}S[\gamma]+...+= \\
  \exp\frac{i}{\hbar}S[\gamma]\biggl(1+\frac{i}{\hbar}\delta S[\gamma]+...+\biggl).
\end{equation*}
Taking into account the terms accurate within the first order of vanishing, we obtain
\begin{equation*}
   \int[d\gamma]\exp\frac{i}{\hbar}S[\gamma]= \int[d\gamma]\exp\frac{i}{\hbar}S[\gamma]+ \int[d\gamma]\frac{i}{\hbar}\delta S[\gamma]\exp\frac{i}{\hbar}S[\gamma]
\end{equation*}
\begin{equation*}
   \int[d\gamma]\delta S[\gamma]\exp\frac{i}{\hbar}S[\gamma]
\end{equation*}
Therefore
\begin{equation*}
  \langle\delta S[\gamma]\rangle=0
\end{equation*}
Using \eqref{eq:math:ex7}, we get
\begin{equation*}
  \langle\delta S[\gamma]\rangle=\biggl\langle \delta\int L(x,\dot{x},\tau)d\tau\biggl\rangle=\biggl(\delta\int L(x_{m},\dot{x_{m}},\tau)d\tau\biggl)\exp\frac{i}{\hbar}S[\gamma_{m}]=\delta S[\gamma_{m}]\exp\frac{i}{\hbar}S[\gamma_{m}].
\end{equation*}
Finally we have
\begin{equation*}
  \delta S[x_{m}(\tau)]=0
\end{equation*}
Thus the individual particles motion obeys classical minimum action principle.

\begin {center} \textbf{\textbf{4. CONCLUSION}}\\
\end {center}
These results can be summarized as follows. If the quantum particle is considered as a matter continuum then description of the motion can be represented in Lagrange's form as mechanical motion of every individual particle along the unique real path. This path is determined by classical minimum action principle.
	This description does not use observables and it is more general then conventional quantum mechanics. Since there are no restrictions to this description, then, probably, it could be used for virtual processes hidden by uncertainty principle.

\vfill\eject

\end{document}